\documentclass[aps,pre,twocolumn,footinbib,superscriptaddress,floatfix]{revtex4-1}
\usepackage{graphicx}
\usepackage{dblfloatfix}
\usepackage{amssymb}
\usepackage{amsmath}
\usepackage{times}
\usepackage{bm}

\begin{document}
			
	\title{Surface super-roughening driven by spatiotemporally correlated noise}	
	\author{Alejandro Al\'es}
	\affiliation{Instituto de F\'isica de Materiales Tandil (IFIMAT), 
		Facultad de Ciencias Exactas, Universidad Nacional 
		del Centro de la Provincia de Buenos Aires (UNCPBA), Pinto 399, 7000 Tandil, Argentina}
	\affiliation{Consejo Nacional de Investigaciones Cient\'ificas y T\'ecnicas (CONICET), Argentina}
	
	\author{Juan M. L\'opez}
	\email{lopez@ifca.unican.es}
	\affiliation{Instituto de F\'isica de Cantabria (IFCA), CSIC-Universidad de
		Cantabria, 39005 Santander, Spain}

	\date{\today}
		
\begin{abstract}
	We study the simple, linear, Edwards-Wilkinson equation that describes surface growth
	governed by height diffusion in the presence of spatiotemporally power-law 
	decaying correlated noise. We analytically show that the surface becomes super-rough 
	when the noise correlations spatio/temporal range is long enough. 
	We calculate analytically the associated anomalous exponents as a function of the
	noise correlation exponents. We also show that super-roughening appears exactly 
	at the threshold point where the local slope surface field becomes rough. In addition, 
	our results indicate that the recent numerical finding of   
	anomalous kinetic roughing of the Kardar-Parisi-Zhang model subject to 
	temporally correlated noise may be inherited from the linear theory.
\end{abstract}	

\maketitle

\section{Introduction}
Surfaces and interfaces driven by random noise often become 
dynamically rough and give rise to 
scale-invariant profiles. Examples can be found in surfaces formed 
by particle deposition processes in thin-film growth ({\it e.g.}, molecular-beam 
epitaxy, sputtering, electrodeposition, and chemical-vapor 
deposition)~\cite{Barabasi.Stanley1995_book,Krug1997}, 
advancing fracture cracks~\cite{Alava2006}, 
and fluid-flow depinning in disordered media~\cite{Alava.etal_2004}, among many others.

A surface $h(\mathbf{x},t)$ is said to be scale-invariant  
if its statistical properties remain unchanged after re-scaling of space and 
time according to the transformation 
$h(\mathbf{x},t) \to b^{\alpha} h(b \, \mathbf{x}, b^{1/z} \, t)$, 
for any scaling factor $b>1$ and a certain choice of critical exponents 
$\alpha$ and $z$~\cite{Barabasi.Stanley1995_book,Krug1997}. 
Scale-invariant surface growth is associated with symmetries~\cite{Hentschel_1994}. 
Indeed, if the dynamical evolution of the surface height $h(\mathbf{x},t)$ 
satisfies a set of fundamental symmetries 
(rotation, translation in $\mathbf{x}$, time invariance, etc), including the 
fundamental shift symmetry $h \to h + c$, where $c$ is a 
constant, then scale-invariant behaviour is expected without 
fine tuning of any external parameters or couplings. 

Scale invariance implies that the local height-height correlations 
exhibit power-law behaviour:
\begin{equation}
\langle \overline{[h(\mathbf{x},t) -
	h(\mathbf{x}+\mathbf{l},t)]^2} \rangle^{1/2} = l^{\alpha} {\cal
	G} (l/t^{1/z}), 
\label{correl}
\end{equation}
where overbar denotes average over all $\mathbf{x}$, brackets
denote average over realizations, and the critical exponents
$\alpha$ and $z$ are the roughening and dynamic exponents,
respectively. The scaling function ${\cal G}(u)$ becomes constant
for $u \ll 1$, and decays as $\sim u^{-\alpha}$ for $u \gg 1$. A similar
scaling describes the global surface width, 
$W(L,t)=\langle\overline{[h(\mathbf{x},t)-\overline{h}(t)]^2}\rangle^{1/2}$,
which is expected to scale as $W(L,t) =  L^{\alpha} {\cal F}
(L/t^{1/z})$, where $\cal F$ is a scaling function with the same asymptotic
behaviour as $\cal G$ in (\ref{correl}).
Therefore, in the stationary regime $t \gg L^z$, we have
$W_\mathrm{stat} (L)\sim L^\alpha$. This scaling picture of kinetically roughened 
surfaces is usually termed Family-Vicsek (FV) ansatz~\cite{Family_1985} and has
shown to be tremendously successful to describe surface growth in a variety of 
theoretical models and experiments~\cite{Barabasi.Stanley1995_book,Krug1997}.

Nowadays, it has become clear that there exist scale-invariant surface
growth models where the above standard FV scaling fails, leading to 
distinctly different scaling functions for the
{\em local} and {\em global} surface fluctuations. Specifically, the local 
height-height correlation function takes the form~(\ref{correl}) 
but with an {\em anomalous} scaling function given 
by~\cite{lopez1997power,lopez1997superroughening}
\begin{equation}
 {\cal G}_\mathrm{A} (u) \sim
 \left\{ \begin{array}{lcl}
  u^{\alpha_\mathrm{loc}-\alpha}   & {\rm if} & u \ll 1 \\
  u^{-\alpha} & {\rm if} & u \gg 1,
 \end{array} 
 \right.
\label{anom_correl}
\end{equation}
for $t \ll L^z$, instead of the standard form (note that standard FV scaling 
is recovered for $\alpha = \alpha_\mathrm{loc}$). 
Therefore, for intermediate 
times, $l^z \ll t \ll L^z$, one has 
$\langle \overline{[h(\mathbf{x},t) -
h(\mathbf{x}+\mathbf{l},t)]^2} \rangle^{1/2} \sim l^{\alpha_\mathrm{loc}} \, t^\kappa$, 
where $\kappa = (\alpha - \alpha_\mathrm{loc})/z$. While the local
surface fluctuations only saturate at times $t \gg L^z$, when they become time independent, 
and one finds $\sim l^{\alpha_\mathrm{loc}} L^{\alpha - \alpha_\mathrm{loc}}$. 
This leads to the existence of an independent {\em local} roughness
exponent $\alpha_\mathrm{loc}$ that characterizes the local interface
fluctuations and differs from the {\em global} roughness exponent
$\alpha$ obtained by, for instance, the global width.   
This phenomenon is referred to as {\em anomalous}
roughening and has received much attention in the last few years
because its commonness in experiments~\cite{ls1998,Morel1998,Myllys2000,Huo.Schwarzacher_2001,Soriano.etal_2002a,Soriano.etal_2005,Auger2006,Planet.etal_2007,Cordoba-Torres.etal_2008,Sana2017,Orrillo12017,Meshkova2018,Planet2018}. 

Current theoretical knowledge has firmly established~\cite{Ramasco.etal_2000} 
that, indeed, the existence of power-law scaling of the correlation functions
({\it i.e.}, scale invariance) does not determine a unique dynamic
scaling form of the correlation functions. Ramasco {\it et al.}~\cite{Ramasco.etal_2000} 
theory of generic dynamic scaling of surface growth predicts the existence of
four possible scaling scenarios. First, the standard FV scaling behavior described
above. Second, there are {\em super-rough} 
processes, $\alpha > 1$, for
which $\alpha_\mathrm{loc} = 1$ always. Third, there are {\em
intrinsically} anomalous roughened surfaces, for which the local
roughness $\alpha_\mathrm{loc} < 1$ is actually an independent exponent
and $\alpha$ may take values larger or smaller than one depending
on the universality class (see~\cite{Lopez2005,Ramasco.etal_2000} and
references therein). Finally, the fourth scaling scenario 
is associated with faceted surfaces, where 
spatio-temporal correlations take a different form, also explained by the
generic dynamic scaling theory of Ramasco {\it et al.}~\cite{Ramasco.etal_2000}. 

The natural question that arises is what kind of
interactions (symmetries, form of the nonlinearities, conservation laws,
non-locality, etc) are required for anomalous roughening to occur in
surface growth?. There are theoretical arguments~\cite{Lopez2005} 
that strongly suggest that local models of surface growth driven by
white noise cannot exhibit intrinsic anomalous roughening, but
super-roughening can occur in models with some conserved dynamics. 
Interestingly, a recent numerical study~\cite{Ales2019}
of the Kardar-Parisi-Zhang (KPZ) equation with long temporally correlated noise 
has found anomalous scaling (associated with facet formation) above some 
critical threshold of the noise correlator index. This indicates
that strong noise correlations may also lead to anomalous roughening.

In this paper we show that, even in the simplest 
local linear growth model, standard FV scaling can break down if the noise 
fluctuations are long-term and/or long-range correlated. We study the 
simplest, linear, scale-invariant growth model, that only takes into
account surface diffusion--- the so-called Edwards-Wilkinson (EW) 
equation~\cite{Edwards1982}--- in the presence 
of long-time and/or long-range correlated noise. 
We show analytically that when spatial/temporal range of the noise 
correlations is long enough the surface becomes super-rough with a 
local roughness exponent $\alpha_\mathrm{loc} = 1$ and a global roughness 
exponent $\alpha > 1$, which value depends on the degree of correlation. 
We also show that super-roughening is associated with the local surface slope
$\nabla h$ becoming rough itself, in agreement with an existing
theoretical conjecture~\cite{Lopez2005}.
We compare our analytical results with the numerical 
integration of the model in the case of temporally correlated noise. 

Our results shed some light into a recent numerical study~\cite{Ales2019}
of the KPZ equation with long-term correlated noise. In that study
it was found that the scaling of the KPZ surface becomes anomalous (and faceted) above some 
critical threshold of the noise correlator index, which was numerically estimated to 
be $\theta_c \approx 0.25$. Here we show that the critical threshold
$\theta_c = 1/4$ appears already in the linear theory as the correlation 
strength above which the surfaces of the linear model become super-rough. 

Theoretical models exhibiting anomalous scaling that are suited for full 
analytical treatment are scarce in the literature, therefore, most examples
have come from either simulations, experiments or scaling approximations. 
Hence, exact analytical results on anomalous behavior are very welcomed. 

\section{Theoretical analysis}
In kinetic surface roughening the EW~\cite{Edwards1982} 
equation plays a central role as the simplest, linear, 
equilibrium model exhibiting scale-invariant behavior. The model
describes surface growth governed by height diffusion (linear elasticity)
subject to noisy thermal-like fluctuations. In $d=1+1$ dimensions it 
can be written as
  \begin{equation}
  \partial_t h(x,t)= \nu \nabla^2 h + \eta(x,t),
  \label{eq:EW1}
  \end{equation}
where the field $h(x,t)$ represents the height of the surface at 
time $t$ and position $x$, $\nu$ is the surface tension, and the noise 
term $\eta (x,t)$ is usually Gaussian and uncorrelated in space and time.

Here we consider the EW equation with long-time and/or long-range correlated, 
instead of thermal uncorrelated noise. For simplicity, we analyse only 
the one-dimensional case, although it is easily
generalizable to larger dimensions. The stochastic term has long-term
correlations with $ \langle \eta (x, t) \rangle = 0 $ and 
\begin{equation}
\langle \eta(x,t) \eta(x',t') \rangle = 2 D |t-t'|^{-(1-2\theta)} |x-x'|^{-(1-2\rho)},
\label{eq:correl1}
\end{equation}
being $D$ the noise amplitude, and $\rho$, $\theta \in [0,1/2)$ are the spatial and 
temporal correlation exponents, respectively, describing the spatial and temporal 
extent of correlations.  This problem was already studied by Pang and Tzeng~\cite{Pang2004}
and the scaling functions were calculated using real space methods. Here we use
a much more transparent calculation of the complete spectral density function (structure factor)
from which the scaling functions and exponents are immediately obtained, including the 
roughness exponent of the local slope field. 

The noise correlator~\eqref{eq:correl1} can be 
written in Fourier space as
 \begin{equation}
 \langle \hat\eta(k,\omega) \hat\eta(k',\omega') \rangle = 
 2D \, |k|^{-2\rho} w^{-2\theta} \delta(k+k')\delta(\omega+\omega').
 \label{eq:correl2}
 \end{equation}
In the limit, $\rho$, $\theta \to 0$ one recovers the case of uncorrelated noise.

In Fourier space the solution of the EW model, 
Eq.~\eqref{eq:EW1}, is given by
\begin{equation}
\hat{h}(k,\omega) = \frac{\hat{\eta}(k,\omega)}{\nu k^2 - i \omega},
\label{eq:EWfft2}
\end{equation}
from which it is possible to obtain the structure factor 
$\Phi(k,\omega) = \langle|\hat{h}(k,\omega)|^2\rangle$
by multiplying the expression~\eqref{eq:EWfft2} by its complex conjugate and 
making a statistical average over the noise; where the correlator~\eqref{eq:correl2} 
of the stochastic term is used. The spectral power density is then given by
\begin{equation}
\Phi(k,\omega) = \frac{2 D k^{-2\rho}\omega^{-2 \theta}}{\nu^2 k^4 + \omega^2}.
\label{eq:phi1}
\end{equation}
By transforming back to real time, inverting the Fourier 
transform in frequencies space, we obtain the exact two-times structure factor 
\begin{equation}
\Phi(k,\Delta  t) = \frac{D}{ \nu^2 2^{-1/2-2\theta}} k^{-2-2\rho-4\theta} f(\nu k^2 \Delta  t),
\label{eq:EWffGnfft1}
\end{equation}
where the scaling function is found to take the form
\begin{align}
f(u) & =     4^{\theta} e^{-u} \nonumber \\ 
&  - u^{1+2\theta} \frac{ {}_1 F_2\{1; 1+\theta,\frac{3}{2}+\theta;\frac{u^2}{4}\} }{\Gamma(1+\theta) \Gamma(\frac{3}{2}+\theta)},
\nonumber  
\end{align}
where $\Gamma$ is the Complete Gamma function and ${}_1 F_2$ the 
Generalized Hypergeometric function~\cite{Olver2010}. The global roughness exponent 
$\alpha$ can be immediately obtained from the power-law 
behavior of Eq.~\eqref{eq:EWffGnfft1}, which has a tail given by
$k^{- \gamma}$, being $ \gamma = 2 \alpha + 1 $. The dynamic 
exponent is given by the argument of the scaling function 
$u=\nu k^2 \Delta t$, which describes a crossover at momenta 
$k = (\nu \Delta t)^{1/z}$, with $z=2$.  
Therefore, the following exact exponents are obtained:
\begin{equation}
\alpha = 1/2 + \rho + 2\, \theta \;\;\; \mathrm{and} \;\;\; z = 2.
\label{eq:EW_exponents}
\end{equation}

Note that the global roughness exponent is greater than unity 
for $\rho + 2 \theta > 1/2$. This already indicates the surface will break
standard FV scaling in this region and we expect to have a different
local roughness exponent $\alpha_\mathrm{loc} \neq \alpha$.
 
In order to calculate $\alpha_\mathrm{loc}$ for this model we resort to the
theory developed in Ref.~\cite{Lopez1999}, which allows us to compute 
the local scaling properties of growth models from the 
dynamics of the local slope field $\Upsilon(x,t) \equiv \nabla h(x,t)$. 

From the EW dynamics in Eq.~\eqref{eq:EW1} we have that the local slope field 
time evolution is given by 
\begin{equation}
\partial_t \Upsilon(x,t) = \nu \nabla^2 \Upsilon(x,t) + \eta^c(x,t),
\nonumber
\end{equation}
where $ \eta^c $ is a correlated and conserved noise with
$$\langle \eta^c(x,t) \eta^c(x',t')\rangle =  
- 2 D\, \nabla^2 |x-x'|^{2\rho-1} |t - t'|^{2\theta-1}.$$
Being a linear equation we can follow the same procedure as before to calculate 
the spectral power density and we have
\begin{equation}
\langle \hat\Upsilon(k,\omega) \hat\Upsilon(-k,-\omega)\rangle \nonumber = 
2 D \,  \frac{k^{2(1-\rho)} \, \omega^{-2\theta} }{\nu^2 k^4 + \omega^2},
\nonumber
\end{equation}
where the extra $k^2$ in the numerator comes from the correlator of the conserved noise, 
but otherwise the formula is the same as Eq.~(\ref{eq:phi1}).
Inverting the Fourier transform in frequencies and by a similar calculation as 
before we arrive at the dynamic exponent $z=2$ (being a linear model, 
both $h$ and $\nabla h$ should have the same dynamic exponent) 
and the roughness exponent
\begin{equation}\hat{\alpha} = -1/2 + 2\theta + \rho,
\label{eq:EWkappa1}
\end{equation}
for the local slope field $\Upsilon(x,t)$. Note that, for $\hat\alpha > 0$, the local slope 
field is a rough surface itself; marking the existence of
anomalous roughening of the original surface $h(x,t)$~\cite{Lopez1999}. 
Accordingly, we expect the local width of $h$ to scale anomalously as 
$w(l,t) \sim l^{\alpha_\mathrm{loc}} \, t^\kappa$ for scales $l^z < t < L^z$ 
when $\hat\alpha > 0$~\cite{Lopez1999}, instead of
being $w(l,t) \sim l^\alpha$ as corresponds to FV scaling. The anomalous time 
exponent corresponds to the time growth of the slope field 
fluctuations $\kappa = \tilde\alpha/z$ and it is also related to the local 
roughness exponent through the expression $\alpha_\mathrm{loc} = \alpha - \kappa z$, 
as described by the general theory in 
Ref.~\cite{Lopez1999}. These two scaling relationships allow us to obtain 
$\alpha_\mathrm{loc}$ from the time growth exponent, $\kappa$, of the 
local slope field. 

Replacing \eqref{eq:EWkappa1} in the expression $\kappa = \hat\alpha/z$, we obtain
\begin{equation}
\kappa = \frac{-1/2 + 2\theta + \rho}{2}
\label{eq:EWkappa2}
\end{equation}
when $\hat\alpha > 0$ and $\kappa = 0$ otherwise.
The local roughness exponent $\alpha_\mathrm{loc}$ follows 
immediately by using~\eqref{eq:EW_exponents} and~\eqref{eq:EWkappa2}

\begin{equation}
\alpha_\mathrm{loc} = \alpha - \kappa z = \left\{
\begin{array}{lcl}
1/2 + 2\theta + \rho \hspace{1.5mm} & \mathrm{if} & 2\theta + \rho \le 1/2  \\
1 & \mathrm{if} & 2\theta + \rho > 1/2
\end{array}
\right.
\label{eq:EWkappa3}     
\end{equation}

Therefore, we find two branches for the value of the local roughness 
exponent $\alpha_\mathrm{loc}$ depending on the noise correlation indexes, 
$\rho$ and $\theta$. On the one hand, the standard FV scaling branch, for
$2\theta + \rho \leq 1/2$, where $\hat\alpha = \kappa = 0$ and we 
have $\alpha = \alpha_\mathrm{loc}=1/2 + 2\theta + \rho$. On the other hand,
the anomalous branch, for $2\theta + \rho > 1/2$, where 
the local slope field becomes rough
($\hat\alpha > 0$) and $\alpha = 1/2 + 2\theta + \rho > 1$ and $\alpha_\mathrm{loc} = 1$, 
which corresponds to super-roughening.

\begin{figure}[h!]
	\begin{center}
		\includegraphics[width=0.35\textwidth]{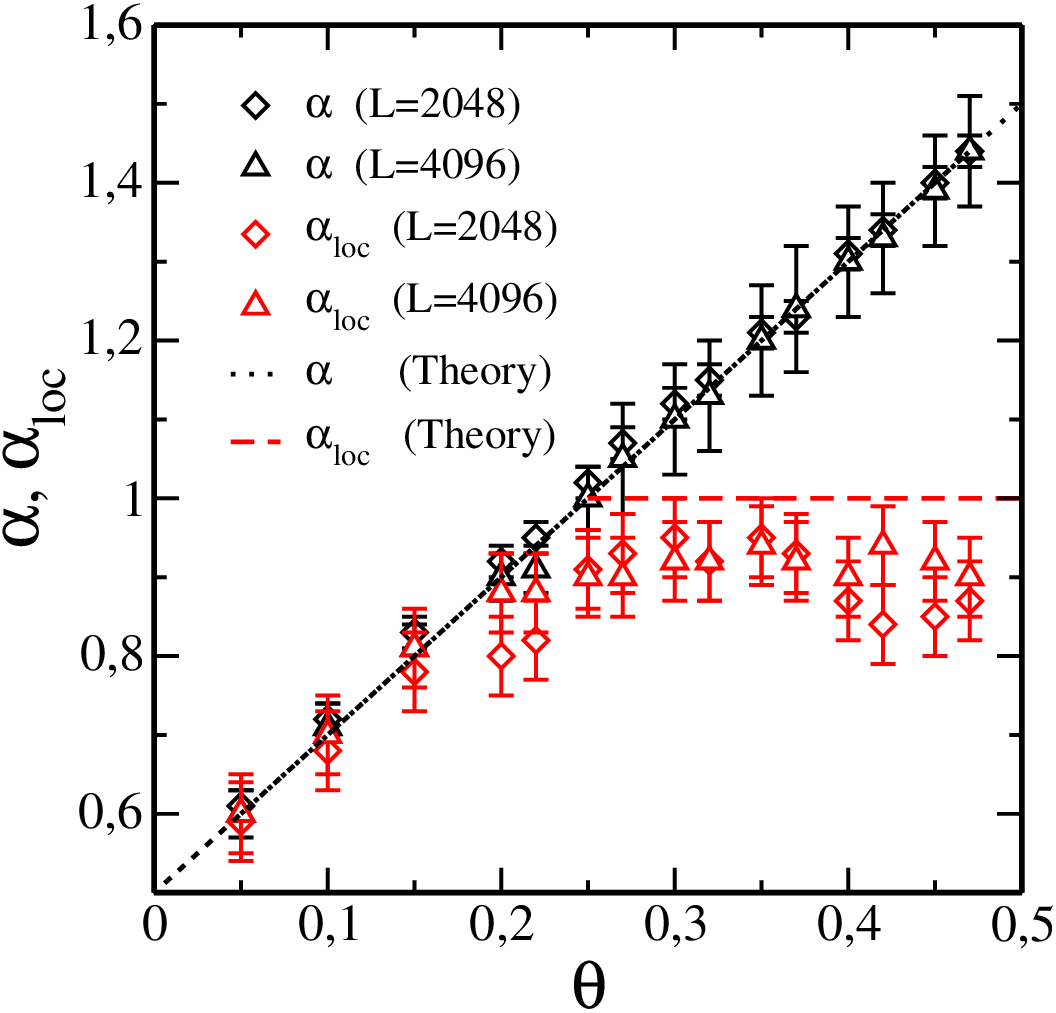}
		\caption{Global $\alpha$ and local $\alpha_\mathrm{loc}$ roughness 
			exponents obtained from numerical integration of the
			EW equation for a temporally correlated noise with index $\theta$. 
			Above the threshold value $\theta_c = 0.25$ 
			the behaviour of the global and local roughness exponents split apart, 
			as predicted by theory, Eqs.~(\ref{eq:EWkappa3}) and~(\ref{eq:EW_exponents}).}		
			\label{fig:alpha-EW}
	\end{center}
\end{figure}
\section{Numerical results}
\label{sct:EWsim}

We now check our analytical results with the numerical 
integration of the EW dynamics with 
long-time correlated noise. We focus on the case of temporally correlated 
noise for which $\rho = 0$ and the index $\theta$ is 
varied in the interval $\theta \in [0,1/2]$. 
Stochastic integration of the EW model was 
carried out by using a standard Euler-Maruyana~\cite{Milstein1994} scheme, 
with a discretization of Eq.~\eqref{eq:EW1} given by
$$h_i(t+\Delta t)= h_i(t) + \Delta t\, L_i(t) + \Delta t\, \eta_i(t),$$
being $ \Delta t $ the temporal step and $ \eta_i (t) $ the noise at position $i$ and time $t$. 
The Laplacian term is discretized as  
$$L_i (t) = \frac {h_{i + 1} (t) + h_{i-1} (t) - 2h_{i} (t)} {\left (\Delta x \right)^2}.$$ 
where lattice constant is set to $ \Delta x = 1 $ and the time 
step $\Delta t = 10^{-3}$. Periodic boundary conditions were used in all our simulations. 
The generation of noise with the desired correlations has been carried out using the Gaussian 
noise fractional technique developed by Mandelbrot~\cite{Mandelbrot1971}, 
which has proven to be very efficient and precise~\cite{Ales2019,Lam1992} for 
systems similar to ours. Statistical averages over 100 independent runs were taken 
for the larger system size and more for smaller ones. We studied system sizes 
$L=1024$, $2048$ and $4096$ and the surface evolution was followed up to times 
of the order $t = 10^6$, for which the system was already in the stationary state.

\begin{figure}[h!]
	\begin{center}
		\includegraphics[width=0.35\textwidth]{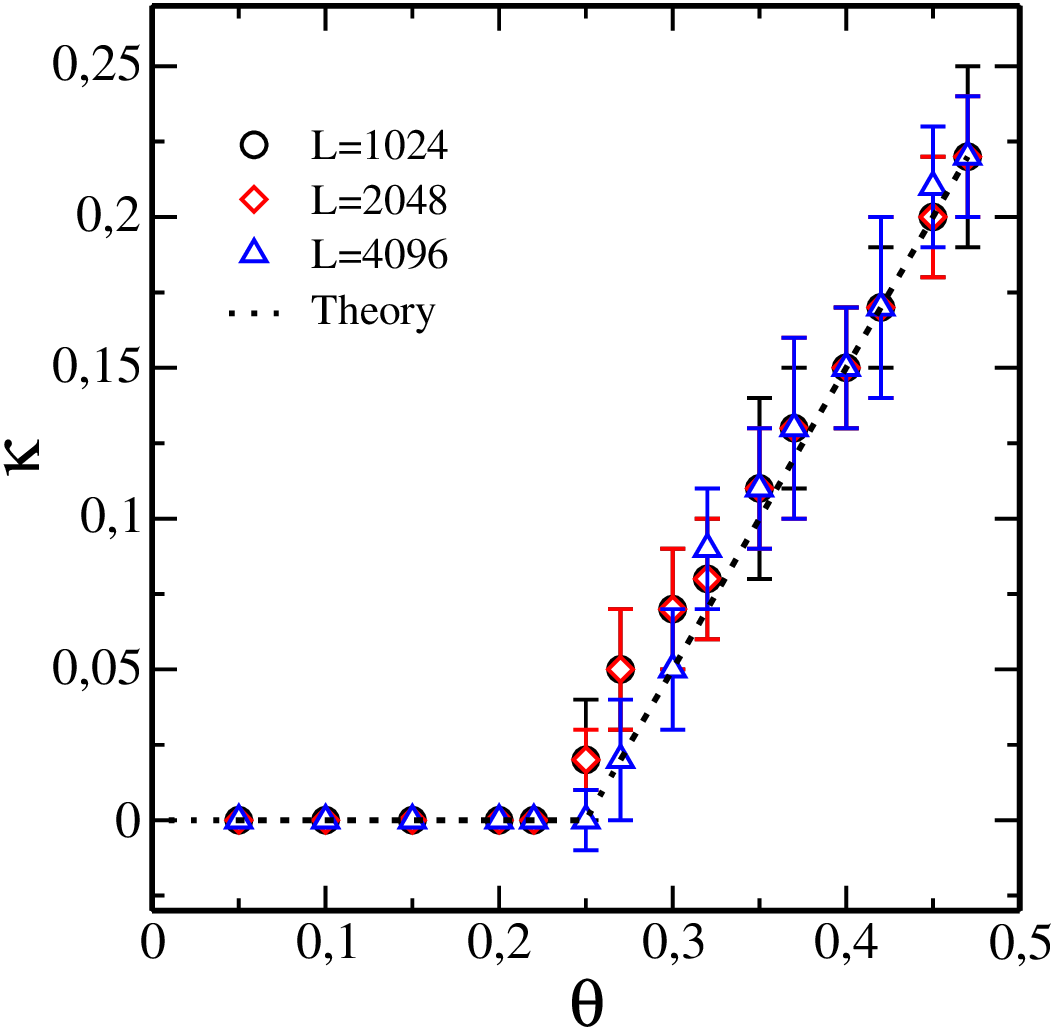}
		\caption{Local growth exponent $ \kappa $ for EW with temporally 
			correlated noise and the analytical prediction, Eq.~(\ref{eq:EWkappa2}). 
			The value of $\kappa$ is positive 
			above $\theta_\mathrm{c} > 0.25$. }
		\label{fig:kappa-EW}
	\end{center}
\end{figure}
\begin{figure}[h!]
	\begin{center}
		\includegraphics[width=0.35\textwidth]{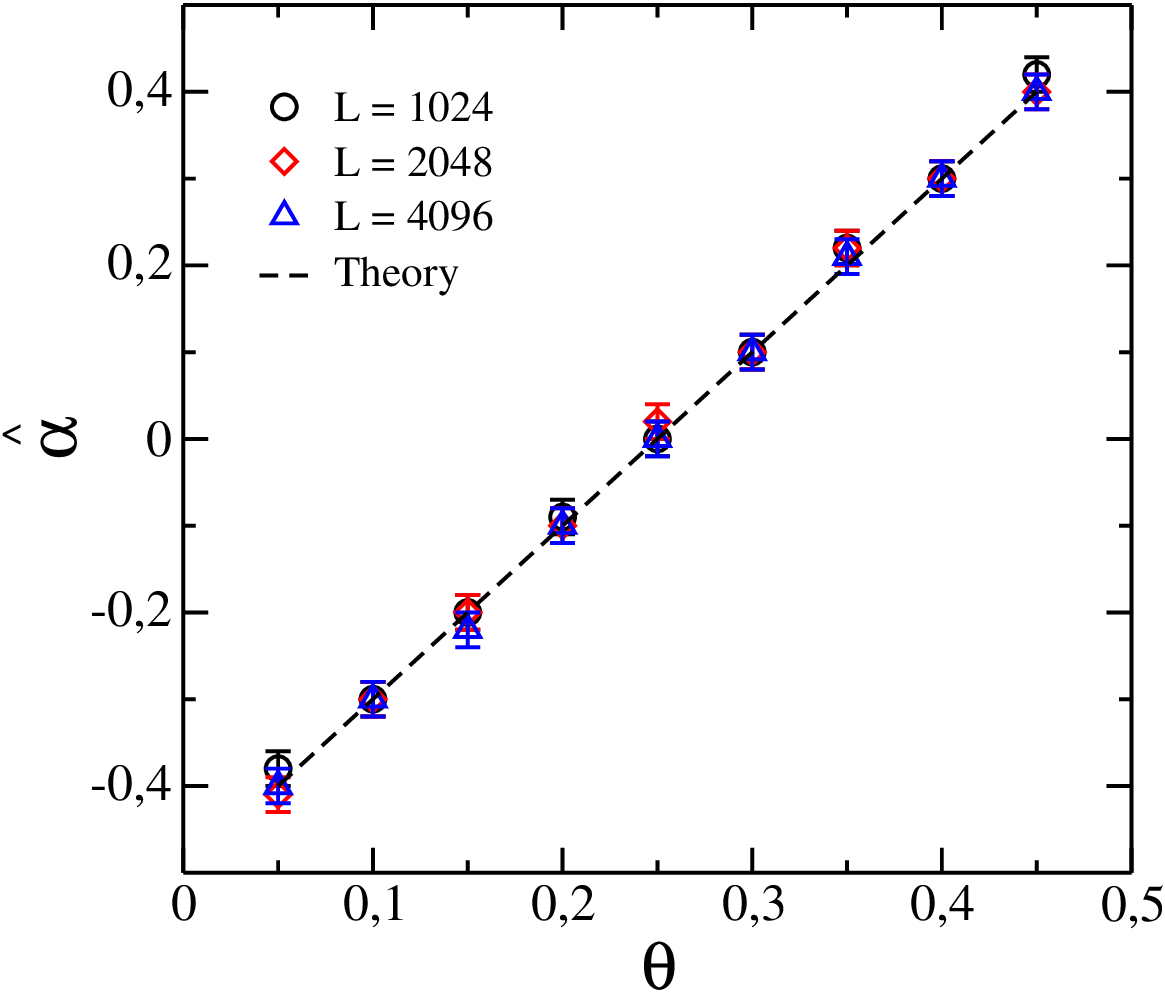}
		\caption{Roughness exponent of the local slope field $\Upsilon(x,t)$ as a function of 
		the noise correlation index $\theta$. Note that $\hat\alpha$ becomes
		positive for $\theta > \theta_c = 1/4$, marking the point above which the slope surface
		field becomes rough.}
		\label{fig:alfa-tilde}
	\end{center}
\end{figure}

We obtained the global roughness exponent numerically from the equal-times spectral density, 
$\langle|\hat h(k,t)|^2\rangle$, for different values of the 
temporal noise correlation exponent $\theta$ within the interval $[0,1/2)$.
We find that, in the stationary 
regime, the spectral density exhibits scaling behaviour 
$\langle|\hat h(k,t \gg L^z)|^2\rangle \sim k^{-(2 \alpha+ 1)}$, 
where $\alpha$ is the global roughness exponent.  
We also computed the local roughness exponent from the data collapse of 
$l^{-\alpha} \langle\overline{[h(\mathbf{x},t) - h(\mathbf{x}+\mathbf{l},t)]^2}\rangle^{1/2}$ 
{\it vs.} $l/t^{1/z}$, 
for a fixed system size $L$, according to the scaling behaviour in 
Eq.~(\ref{anom_correl}). This gives us independent measurements of $z$, $\alpha$, 
and produces an estimate of $\alpha_\mathrm{loc}$ from the power fit of the scaling
function asymptote, $\sim (l/t^{1/z})^{2(\alpha_\mathrm{loc} - \alpha)}$, for $l/t^{1/z} \ll 1$.
These results are summarized in Fig.~\ref{fig:alpha-EW}, where one can observe 
that $\alpha = \alpha_\mathrm{loc}$ for a correlation index 
$\theta < 1/4$, while the roughness exponents 
split above that threshold when $\alpha_\mathrm{loc}$ gets values around $0.9$.
The figure also shows the
theoretical prediction given by Eqs.~\eqref{eq:EW_exponents} and~\eqref{eq:EWkappa3}.
   
We also measured the anomalous time exponent $\kappa$ from the time-growth of the 
local slope fluctuations, $\langle\overline{(\nabla h)^2}\rangle \sim t^{2\kappa}$. 
Alternatively, $\kappa$ could also be measured from the time behaviour of the height-height
correlation at a fixed scale $l \ll L$, 
$\overline{[h(\mathbf{x},t) - h(\mathbf{x}+\mathbf{l},t \gg L^z)]^2} \rangle 
\sim l^{2\alpha_\mathrm{loc}} \, t^{2\kappa}$~\cite{Ramasco.etal_2000}. 
In Fig.~\ref{fig:kappa-EW} we show 
our results for $\kappa$ as a function of $\theta$ together with the analytical 
prediction. We observe that, in agreement with the theoretical
prediction, $\kappa = 0$ for $\theta < 1/4$ and follows 
Eq.~\eqref{eq:EWkappa2} above that point.

Finally, we measured the roughness exponent of the local slope field $\Upsilon(x,t)$ by
computing the spectral density in the stationary state,  
$\langle|\hat \Upsilon(k,t \gg L^z)|^2\rangle \sim k^{-(2\hat\alpha + 1)}$, 
from here we obtained the roughness
exponent $\hat\alpha$ for the slope field. Figure~\ref{fig:alfa-tilde} 
summarizes our numerical results
as the noise correlation index $\theta$ is varied. 

\section{Conclusions}

We have studied the simple, linear, EW dynamics of surface growth in
the presence of noise with spatio-temporal power-law 
decaying correlations. We have found analytically 
that the surface becomes super-rough ($\alpha > 1$ and $\alpha_\mathrm{loc} = 1$)
for correlation noise indexes satisfying $\rho + 2\theta > 1/2$.
Our calculation in Fourier space is more transparent and 
much simpler than an earlier method~\cite{Pang2004} based on a direct
calculation of the height-height correlation in real space.    
Taking advantage of the simplicity of the calculation in Fourier space, 
we also derived an exact prediction for the anomalous time exponent $\kappa$
as a function of the noise correlation 
by resorting to studying the dynamics of the local slope surface. This technique
allowed us to study the roughness exponent $\hat\alpha$ of the surface slope field 
and show analytically that the appearance of the super-roughening 
regime of the EW surface is associated with the slope field becoming 
rough itself, {\it i.e.} $\hat\alpha > 0$, in agreement with 
existing theory~\cite{Lopez2005}. In addition, we compared 
our theoretical predictions with a numerical integration of the 
EW equation in the case of temporally correlated noise.

In the particular case of temporal correlated noise ($\rho = 0$), 
our results lead to the exact threshold $\theta_\mathrm{c} = 1/4$, 
which is very close, if not identical, to the critical value recently found numerically 
for the appearance of anomalous scaling in the nonlinear KPZ equation~\cite{Ales2019}.
It is remarkable, and not fully understood yet, that the critical threshold
for anomalous kinetic roughening in KPZ with temporally correlated noise 
appears to be inherited from the linear theory. 

\section*{Acknowledgments}

This work has been partially supported by the Program for 
Scientific Cooperation “I-COOP+” from Consejo Superior de 
Investigaciones Cient\'ificas (Spain) through project No. COOPA20187.
A.\ A. is grateful for the financial support from Programa de 
Pasant\'ias de la Universidad de Cantabria in 2017 and 2018 
(Projects No. 70-ZCE3-226.90 and 62-VCES-648), and CONICET 
(Argentina) for a post-doctoral fellowship. 
J.\ M.\ L is partially supported by project No. FIS2016-74957-P
from Agencia Estatal de Investigaci\'on (AEI) and FEDER (EU).


\begin{thebibliography}{31}%
	\makeatletter
	\providecommand \@ifxundefined [1]{%
		\@ifx{#1\undefined}
	}%
	\providecommand \@ifnum [1]{%
		\ifnum #1\expandafter \@firstoftwo
		\else \expandafter \@secondoftwo
		\fi
	}%
	\providecommand \@ifx [1]{%
		\ifx #1\expandafter \@firstoftwo
		\else \expandafter \@secondoftwo
		\fi
	}%
	\providecommand \natexlab [1]{#1}%
	\providecommand \enquote  [1]{``#1''}%
	\providecommand \bibnamefont  [1]{#1}%
	\providecommand \bibfnamefont [1]{#1}%
	\providecommand \citenamefont [1]{#1}%
	\providecommand \href@noop [0]{\@secondoftwo}%
	\providecommand \href [0]{\begingroup \@sanitize@url \@href}%
	\providecommand \@href[1]{\@@startlink{#1}\@@href}%
	\providecommand \@@href[1]{\endgroup#1\@@endlink}%
	\providecommand \@sanitize@url [0]{\catcode `\\12\catcode `\$12\catcode
		`\&12\catcode `\#12\catcode `\^12\catcode `\_12\catcode `\%12\relax}%
	\providecommand \@@startlink[1]{}%
	\providecommand \@@endlink[0]{}%
	\providecommand \url  [0]{\begingroup\@sanitize@url \@url }%
	\providecommand \@url [1]{\endgroup\@href {#1}{\urlprefix }}%
	\providecommand \urlprefix  [0]{URL }%
	\providecommand \Eprint [0]{\href }%
	\providecommand \doibase [0]{http://dx.doi.org/}%
	\providecommand \selectlanguage [0]{\@gobble}%
	\providecommand \bibinfo  [0]{\@secondoftwo}%
	\providecommand \bibfield  [0]{\@secondoftwo}%
	\providecommand \translation [1]{[#1]}%
	\providecommand \BibitemOpen [0]{}%
	\providecommand \bibitemStop [0]{}%
	\providecommand \bibitemNoStop [0]{.\EOS\space}%
	\providecommand \EOS [0]{\spacefactor3000\relax}%
	\providecommand \BibitemShut  [1]{\csname bibitem#1\endcsname}%
	\let\auto@bib@innerbib\@empty
	\bibitem [{\citenamefont {Barab{\'a}si}\ and\ \citenamefont
		{Stanley}(1995)}]{Barabasi.Stanley1995_book}%
	\BibitemOpen
	\bibfield  {author} {\bibinfo {author} {\bibfnamefont {A.-L.}\ \bibnamefont
			{Barab{\'a}si}}\ and\ \bibinfo {author} {\bibfnamefont {H.~E.}\ \bibnamefont
			{Stanley}},\ }\href@noop {} {\emph {\bibinfo {title} {Fractal Concepts in
				Surface Growth}}}\ (\bibinfo  {publisher} {Cambridge University Press},\
	\bibinfo {address} {Cambridge},\ \bibinfo {year} {1995})\
	\BibitemShut {NoStop}%
	\bibitem [{\citenamefont {Krug}(1997)}]{Krug1997}%
	\BibitemOpen
	\bibfield  {author} {\bibinfo {author} {\bibfnamefont {J.}~\bibnamefont
			{Krug}},\ }\href {\doibase 10.1080/00018739700101498} {\bibfield  {journal}
		{\bibinfo  {journal} {Adv. Phys.}\ }\textbf {\bibinfo {volume} {46}},\
		\bibinfo {pages} {139} (\bibinfo {year} {1997})}\BibitemShut {NoStop}%
	\bibitem [{\citenamefont {Alava}\ \emph {et~al.}(2006)\citenamefont {Alava},
		\citenamefont {Nukala},\ and\ \citenamefont {Zapperi}}]{Alava2006}%
	\BibitemOpen
	\bibfield  {author} {\bibinfo {author} {\bibfnamefont {M.~J.}\ \bibnamefont
			{Alava}}, \bibinfo {author} {\bibfnamefont {P.~K. V.~V.}\ \bibnamefont
			{Nukala}}, \ and\ \bibinfo {author} {\bibfnamefont {S.}~\bibnamefont
			{Zapperi}},\ }\href {\doibase 10.1080/00018730300741518} {\bibfield
		{journal} {\bibinfo  {journal} {Adv. Phys.}\ }\textbf {\bibinfo {volume}
			{55}},\ \bibinfo {pages} {349} (\bibinfo {year} {2006})}\BibitemShut
	{NoStop}%
	\bibitem [{\citenamefont {Alava}\ \emph {et~al.}(2004)\citenamefont {Alava},
		\citenamefont {Dub{\'e}},\ and\ \citenamefont {Rost}}]{Alava.etal_2004}%
	\BibitemOpen
	\bibfield  {author} {\bibinfo {author} {\bibfnamefont {M.}~\bibnamefont
			{Alava}}, \bibinfo {author} {\bibfnamefont {M.}~\bibnamefont {Dub{\'e}}}, \
		and\ \bibinfo {author} {\bibfnamefont {M.}~\bibnamefont {Rost}},\ }\href@noop
	{} {\bibfield  {journal} {\bibinfo  {journal} {Adv. Phys.}\ }\textbf
		{\bibinfo {volume} {53}},\ \bibinfo {pages} {83} (\bibinfo {year}
		{2004})}\BibitemShut {NoStop}%
	\bibitem [{\citenamefont {Hentschel}(1994)}]{Hentschel_1994}%
	\BibitemOpen
	\bibfield  {author} {\bibinfo {author} {\bibfnamefont {H.~G.~E.}\
			\bibnamefont {Hentschel}},\ }\href {\doibase 10.1088/0305-4470/27/7/008}
	{\bibfield  {journal} {\bibinfo  {journal} {J. Phys. A: Math. Gen.}\ }\textbf
		{\bibinfo {volume} {27}},\ \bibinfo {pages} {2269} (\bibinfo {year}
		{1994})}\BibitemShut {NoStop}%
	\bibitem [{\citenamefont {Family}\ and\ \citenamefont
		{Vicsek}(1985)}]{Family_1985}%
	\BibitemOpen
	\bibfield  {author} {\bibinfo {author} {\bibfnamefont {F.}~\bibnamefont
			{Family}}\ and\ \bibinfo {author} {\bibfnamefont {T.}~\bibnamefont
			{Vicsek}},\ }\href {\doibase 10.1088/0305-4470/18/2/005} {\bibfield
		{journal} {\bibinfo  {journal} {J. Phys. A: Math. Gen.}\ }\textbf {\bibinfo
			{volume} {18}},\ \bibinfo {pages} {L75} (\bibinfo {year} {1985})}\BibitemShut
	{NoStop}%
	\bibitem [{\citenamefont {L{\'o}pez}\ \emph
		{et~al.}(1997{\natexlab{a}})\citenamefont {L{\'o}pez}, \citenamefont
		{Rodr{\'\i}guez},\ and\ \citenamefont {Cuerno}}]{lopez1997power}%
	\BibitemOpen
	\bibfield  {author} {\bibinfo {author} {\bibfnamefont {J.~M.}\ \bibnamefont
			{L{\'o}pez}}, \bibinfo {author} {\bibfnamefont {M.~A.}\ \bibnamefont
			{Rodr{\'\i}guez}}, \ and\ \bibinfo {author} {\bibfnamefont {R.}~\bibnamefont
			{Cuerno}},\ }\href@noop {} {\bibfield  {journal} {\bibinfo  {journal}
			{Physica A}\ }\textbf {\bibinfo {volume} {246}},\ \bibinfo {pages} {329}
		(\bibinfo {year} {1997}{\natexlab{a}})}\BibitemShut {NoStop}%
	\bibitem [{\citenamefont {L{\'o}pez}\ \emph
		{et~al.}(1997{\natexlab{b}})\citenamefont {L{\'o}pez}, \citenamefont
		{Rodriguez},\ and\ \citenamefont {Cuerno}}]{lopez1997superroughening}%
	\BibitemOpen
	\bibfield  {author} {\bibinfo {author} {\bibfnamefont {J.~M.}\ \bibnamefont
			{L{\'o}pez}}, \bibinfo {author} {\bibfnamefont {M.~A.}\ \bibnamefont
			{Rodriguez}}, \ and\ \bibinfo {author} {\bibfnamefont {R.}~\bibnamefont
			{Cuerno}},\ }\href@noop {} {\bibfield  {journal} {\bibinfo  {journal} {Phys.
				Rev. E}\ }\textbf {\bibinfo {volume} {56}},\ \bibinfo {pages} {3993}
		(\bibinfo {year} {1997}{\natexlab{b}})}\BibitemShut {NoStop}%
	\bibitem [{\citenamefont {L{\'o}pez}\ and\ \citenamefont
		{Schmittbuhl}(1998)}]{ls1998}%
	\BibitemOpen
	\bibfield  {author} {\bibinfo {author} {\bibfnamefont {J.~M.}\ \bibnamefont
			{L{\'o}pez}}\ and\ \bibinfo {author} {\bibfnamefont {J.}~\bibnamefont
			{Schmittbuhl}},\ }\href@noop {} {\bibfield  {journal} {\bibinfo  {journal}
			{Phys. Rev. E}\ }\textbf {\bibinfo {volume} {57}},\ \bibinfo {pages} {6405}
		(\bibinfo {year} {1998})}\BibitemShut {NoStop}%
	\bibitem [{\citenamefont {Morel}\ \emph {et~al.}(1998)\citenamefont {Morel},
		\citenamefont {Schmittbuhl}, \citenamefont {L\'{o}pez},\ and\ \citenamefont
		{Valentin}}]{Morel1998}%
	\BibitemOpen
	\bibfield  {author} {\bibinfo {author} {\bibfnamefont {S.}~\bibnamefont
			{Morel}}, \bibinfo {author} {\bibfnamefont {J.}~\bibnamefont {Schmittbuhl}},
		\bibinfo {author} {\bibfnamefont {J.~M.}\ \bibnamefont {L\'{o}pez}}, \ and\
		\bibinfo {author} {\bibfnamefont {G.}~\bibnamefont {Valentin}},\ }\href
	{http://link.aps.org/abstract/PRE/v58/p6999} {\bibfield  {journal} {\bibinfo
			{journal} {Phys. Rev. E}\ }\textbf {\bibinfo {volume} {58}},\ \bibinfo
		{pages} {6999} (\bibinfo {year} {1998})}\BibitemShut {NoStop}%
	\bibitem [{\citenamefont {Myllys}\ \emph {et~al.}(2000)\citenamefont {Myllys},
		\citenamefont {Maunuksela}, \citenamefont {Alava}, \citenamefont
		{Ala-Nissila},\ and\ \citenamefont {Timonen}}]{Myllys2000}%
	\BibitemOpen
	\bibfield  {author} {\bibinfo {author} {\bibfnamefont {M.}~\bibnamefont
			{Myllys}}, \bibinfo {author} {\bibfnamefont {J.}~\bibnamefont {Maunuksela}},
		\bibinfo {author} {\bibfnamefont {M.~J.}\ \bibnamefont {Alava}}, \bibinfo
		{author} {\bibfnamefont {T.}~\bibnamefont {Ala-Nissila}}, \ and\ \bibinfo
		{author} {\bibfnamefont {J.}~\bibnamefont {Timonen}},\ }\href {\doibase
		10.1103/PhysRevLett.84.1946} {\bibfield  {journal} {\bibinfo  {journal}
			{Phys. Rev. Lett.}\ }\textbf {\bibinfo {volume} {84}},\ \bibinfo {pages}
		{1946} (\bibinfo {year} {2000})}\BibitemShut {NoStop}%
	\bibitem [{\citenamefont {Huo}\ and\ \citenamefont
		{Schwarzacher}(2001)}]{Huo.Schwarzacher_2001}%
	\BibitemOpen
	\bibfield  {author} {\bibinfo {author} {\bibfnamefont {S.}~\bibnamefont
			{Huo}}\ and\ \bibinfo {author} {\bibfnamefont {W.}~\bibnamefont
			{Schwarzacher}},\ }\href {http://link.aps.org/abstract/PRL/v86/p256}
	{\bibfield  {journal} {\bibinfo  {journal} {Phys. Rev. Lett.}\ }\textbf
		{\bibinfo {volume} {86}},\ \bibinfo {pages} {256} (\bibinfo {year}
		{2001})}\BibitemShut {NoStop}%
	\bibitem [{\citenamefont {Soriano}\ \emph {et~al.}(2002)\citenamefont
		{Soriano}, \citenamefont {Ramasco}, \citenamefont {Rodr\'{\i}guez},
		\citenamefont {Hern{\'a}ndez-Machado},\ and\ \citenamefont
		{Ort\'{\i}n}}]{Soriano.etal_2002a}%
	\BibitemOpen
	\bibfield  {author} {\bibinfo {author} {\bibfnamefont {J.}~\bibnamefont
			{Soriano}}, \bibinfo {author} {\bibfnamefont {J.~J.}\ \bibnamefont
			{Ramasco}}, \bibinfo {author} {\bibfnamefont {M.~A.}\ \bibnamefont
			{Rodr\'{\i}guez}}, \bibinfo {author} {\bibfnamefont {A.}~\bibnamefont
			{Hern{\'a}ndez-Machado}}, \ and\ \bibinfo {author} {\bibfnamefont
			{J.}~\bibnamefont {Ort\'{\i}n}},\ }\href@noop {} {\bibfield  {journal}
		{\bibinfo  {journal} {Phys. Rev. Lett.}\ }\textbf {\bibinfo {volume} {89}},\
		\bibinfo {pages} {026102} (\bibinfo {year} {2002})}\BibitemShut {NoStop}%
	\bibitem [{\citenamefont {Soriano}\ \emph {et~al.}(2005)\citenamefont
		{Soriano}, \citenamefont {Mercier}, \citenamefont {Planet}, \citenamefont
		{Hern{\'a}ndez-Machado}, \citenamefont {Rodr{\'\i}guez},\ and\ \citenamefont
		{Ort{\'\i}n}}]{Soriano.etal_2005}%
	\BibitemOpen
	\bibfield  {author} {\bibinfo {author} {\bibfnamefont {J.}~\bibnamefont
			{Soriano}}, \bibinfo {author} {\bibfnamefont {A.}~\bibnamefont {Mercier}},
		\bibinfo {author} {\bibfnamefont {R.}~\bibnamefont {Planet}}, \bibinfo
		{author} {\bibfnamefont {A.}~\bibnamefont {Hern{\'a}ndez-Machado}}, \bibinfo
		{author} {\bibfnamefont {M.~A.}\ \bibnamefont {Rodr{\'\i}guez}}, \ and\
		\bibinfo {author} {\bibfnamefont {J.}~\bibnamefont {Ort{\'\i}n}},\
	}\href@noop {} {\bibfield  {journal} {\bibinfo  {journal} {Phys. Rev. Lett.}\
		}\textbf {\bibinfo {volume} {95}},\ \bibinfo {pages} {104501} (\bibinfo
		{year} {2005})}\BibitemShut {NoStop}%
	\bibitem [{\citenamefont {Auger}\ \emph {et~al.}(2006)\citenamefont {Auger},
		\citenamefont {V\'azquez}, \citenamefont {Cuerno}, \citenamefont {Castro},
		\citenamefont {Jergel},\ and\ \citenamefont {S\'anchez}}]{Auger2006}%
	\BibitemOpen
	\bibfield  {author} {\bibinfo {author} {\bibfnamefont {M.~A.}\ \bibnamefont
			{Auger}}, \bibinfo {author} {\bibfnamefont {L.}~\bibnamefont {V\'azquez}},
		\bibinfo {author} {\bibfnamefont {R.}~\bibnamefont {Cuerno}}, \bibinfo
		{author} {\bibfnamefont {M.}~\bibnamefont {Castro}}, \bibinfo {author}
		{\bibfnamefont {M.}~\bibnamefont {Jergel}}, \ and\ \bibinfo {author}
		{\bibfnamefont {O.}~\bibnamefont {S\'anchez}},\ }\href {\doibase
		10.1103/PhysRevB.73.045436} {\bibfield  {journal} {\bibinfo  {journal} {Phys.
				Rev. B}\ }\textbf {\bibinfo {volume} {73}},\ \bibinfo {pages} {045436}
		(\bibinfo {year} {2006})}\BibitemShut {NoStop}%
	\bibitem [{\citenamefont {Planet}\ \emph {et~al.}(2007)\citenamefont {Planet},
		\citenamefont {Pradas}, \citenamefont {Hern{\'a}ndez-Machado},\ and\
		\citenamefont {Ort\'{\i}n}}]{Planet.etal_2007}%
	\BibitemOpen
	\bibfield  {author} {\bibinfo {author} {\bibfnamefont {R.}~\bibnamefont
			{Planet}}, \bibinfo {author} {\bibfnamefont {M.}~\bibnamefont {Pradas}},
		\bibinfo {author} {\bibfnamefont {A.}~\bibnamefont {Hern{\'a}ndez-Machado}},
		\ and\ \bibinfo {author} {\bibfnamefont {J.}~\bibnamefont {Ort\'{\i}n}},\
	}\href@noop {} {\bibfield  {journal} {\bibinfo  {journal} {Phys. Rev. E}\
		}\textbf {\bibinfo {volume} {76}},\ \bibinfo {pages} {056312} (\bibinfo
		{year} {2007})}\BibitemShut {NoStop}%
	\bibitem [{\citenamefont {Cordoba-Torres}\ \emph {et~al.}(2008)\citenamefont
		{Cordoba-Torres}, \citenamefont {Bastos},\ and\ \citenamefont
		{Nogueira}}]{Cordoba-Torres.etal_2008}%
	\BibitemOpen
	\bibfield  {author} {\bibinfo {author} {\bibfnamefont {P.}~\bibnamefont
			{Cordoba-Torres}}, \bibinfo {author} {\bibfnamefont {I.~N.}\ \bibnamefont
			{Bastos}}, \ and\ \bibinfo {author} {\bibfnamefont {R.~P.}\ \bibnamefont
			{Nogueira}},\ }\href {\doibase 10.1103/PhysRevE.77.031602} {\bibfield
		{journal} {\bibinfo  {journal} {Phys. Rev. E}\ }\textbf {\bibinfo {volume}
			{77}},\ \bibinfo {eid} {031602} (\bibinfo {year} {2008})}\BibitemShut
	{NoStop}%
	\bibitem [{\citenamefont {Sana}\ \emph {et~al.}(2017)\citenamefont {Sana},
		\citenamefont {V{\'{a}}zquez}, \citenamefont {Cuerno},\ and\ \citenamefont
		{Sarkar}}]{Sana2017}%
	\BibitemOpen
	\bibfield  {author} {\bibinfo {author} {\bibfnamefont {P.}~\bibnamefont
			{Sana}}, \bibinfo {author} {\bibfnamefont {L.}~\bibnamefont {V{\'{a}}zquez}},
		\bibinfo {author} {\bibfnamefont {R.}~\bibnamefont {Cuerno}}, \ and\ \bibinfo
		{author} {\bibfnamefont {S.}~\bibnamefont {Sarkar}},\ }\href {\doibase
		10.1088/1361-6463/aa87e8} {\bibfield  {journal} {\bibinfo  {journal} {J.
				Phys. D: Appl. Phys.}\ }\textbf {\bibinfo {volume} {50}},\ \bibinfo {pages}
		{435306} (\bibinfo {year} {2017})}\BibitemShut {NoStop}%
	\bibitem [{\citenamefont {Orrillo}\ \emph {et~al.}(2017)\citenamefont
		{Orrillo}, \citenamefont {Santalla}, \citenamefont {Cuerno}, \citenamefont
		{Vázquez}, \citenamefont {Ribotta}, \citenamefont {Gassa}, \citenamefont
		{Mompean}, \citenamefont {Salvarezza},\ and\ \citenamefont
		{Vela}}]{Orrillo12017}%
	\BibitemOpen
	\bibfield  {author} {\bibinfo {author} {\bibfnamefont {P.~A.}\ \bibnamefont
			{Orrillo}}, \bibinfo {author} {\bibfnamefont {S.~N.}\ \bibnamefont
			{Santalla}}, \bibinfo {author} {\bibfnamefont {R.}~\bibnamefont {Cuerno}},
		\bibinfo {author} {\bibfnamefont {L.}~\bibnamefont {Vázquez}}, \bibinfo
		{author} {\bibfnamefont {S.~B.}\ \bibnamefont {Ribotta}}, \bibinfo {author}
		{\bibfnamefont {L.~M.}\ \bibnamefont {Gassa}}, \bibinfo {author}
		{\bibfnamefont {F.~J.}\ \bibnamefont {Mompean}}, \bibinfo {author}
		{\bibfnamefont {R.~C.}\ \bibnamefont {Salvarezza}}, \ and\ \bibinfo {author}
		{\bibfnamefont {M.~E.}\ \bibnamefont {Vela}},\ }\href {\doibase
		10.1038/s41598-017-18155-7} {\bibfield  {journal} {\bibinfo  {journal} {Sci.
				Rep.}\ }\textbf {\bibinfo {volume} {7}},\ \bibinfo {pages} {17997} (\bibinfo
		{year} {2017})}\BibitemShut {NoStop}%
	\bibitem [{\citenamefont {Meshkova}\ \emph {et~al.}(2018)\citenamefont
		{Meshkova}, \citenamefont {Starostin}, \citenamefont {van~de Sanden},\ and\
		\citenamefont {de~Vries}}]{Meshkova2018}%
	\BibitemOpen
	\bibfield  {author} {\bibinfo {author} {\bibfnamefont {A.~S.}\ \bibnamefont
			{Meshkova}}, \bibinfo {author} {\bibfnamefont {S.~A.}\ \bibnamefont
			{Starostin}}, \bibinfo {author} {\bibfnamefont {M.~C.~M.}\ \bibnamefont
			{van~de Sanden}}, \ and\ \bibinfo {author} {\bibfnamefont {H.~W.}\
			\bibnamefont {de~Vries}},\ }\href {\doibase 10.1088/1361-6463/aacb1c}
	{\bibfield  {journal} {\bibinfo  {journal} {J. Phys. D: Appl. Phys.}\
		}\textbf {\bibinfo {volume} {51}},\ \bibinfo {pages} {285303} (\bibinfo
		{year} {2018})}\BibitemShut {NoStop}%
	\bibitem [{\citenamefont {Planet}\ \emph {et~al.}(2018)\citenamefont {Planet},
		\citenamefont {L\'opez}, \citenamefont {Santucci},\ and\ \citenamefont
		{Ort\'{\i}n}}]{Planet2018}%
	\BibitemOpen
	\bibfield  {author} {\bibinfo {author} {\bibfnamefont {R.}~\bibnamefont
			{Planet}}, \bibinfo {author} {\bibfnamefont {J.~M.}\ \bibnamefont {L\'opez}},
		\bibinfo {author} {\bibfnamefont {S.}~\bibnamefont {Santucci}}, \ and\
		\bibinfo {author} {\bibfnamefont {J.}~\bibnamefont {Ort\'{\i}n}},\ }\href
	{\doibase 10.1103/PhysRevLett.121.034101} {\bibfield  {journal} {\bibinfo
			{journal} {Phys. Rev. Lett.}\ }\textbf {\bibinfo {volume} {121}},\ \bibinfo
		{pages} {034101} (\bibinfo {year} {2018})}\BibitemShut {NoStop}%
	\bibitem [{\citenamefont {Ramasco}\ \emph {et~al.}(2000)\citenamefont
		{Ramasco}, \citenamefont {L\'{o}pez},\ and\ \citenamefont
		{Rodr\'{i}guez}}]{Ramasco.etal_2000}%
	\BibitemOpen
	\bibfield  {author} {\bibinfo {author} {\bibfnamefont {J.~J.}\ \bibnamefont
			{Ramasco}}, \bibinfo {author} {\bibfnamefont {J.~M.}\ \bibnamefont
			{L\'{o}pez}}, \ and\ \bibinfo {author} {\bibfnamefont {M.~A.}\ \bibnamefont
			{Rodr\'{i}guez}},\ }\href {http://link.aps.org/abstract/PRL/v84/p2199}
	{\bibfield  {journal} {\bibinfo  {journal} {Phys. Rev. Lett.}\ }\textbf
		{\bibinfo {volume} {84}},\ \bibinfo {pages} {2199} (\bibinfo {year}
		{2000})}\BibitemShut {NoStop}%
	\bibitem [{\citenamefont {L{\'o}pez}\ \emph {et~al.}(2005)\citenamefont
		{L{\'o}pez}, \citenamefont {Castro},\ and\ \citenamefont
		{Gallego}}]{Lopez2005}%
	\BibitemOpen
	\bibfield  {author} {\bibinfo {author} {\bibfnamefont {J.~M.}\ \bibnamefont
			{L{\'o}pez}}, \bibinfo {author} {\bibfnamefont {M.}~\bibnamefont {Castro}}, \
		and\ \bibinfo {author} {\bibfnamefont {R.}~\bibnamefont {Gallego}},\
	}\href@noop {} {\bibfield  {journal} {\bibinfo  {journal} {Phys. Rev. Lett.}\
		}\textbf {\bibinfo {volume} {94}},\ \bibinfo {pages} {166103} (\bibinfo
		{year} {2005})}\BibitemShut {NoStop}%
	\bibitem [{\citenamefont {Al\'es}\ and\ \citenamefont
		{L\'opez}(2019)}]{Ales2019}%
	\BibitemOpen
	\bibfield  {author} {\bibinfo {author} {\bibfnamefont {A.}~\bibnamefont
			{Al\'es}}\ and\ \bibinfo {author} {\bibfnamefont {J.~M.}\ \bibnamefont
			{L\'opez}},\ }\href {\doibase 10.1103/PhysRevE.99.062139} {\bibfield
		{journal} {\bibinfo  {journal} {Phys. Rev. E}\ }\textbf {\bibinfo {volume}
			{99}},\ \bibinfo {pages} {062139} (\bibinfo {year} {2019})}\BibitemShut
	{NoStop}%
	\bibitem [{\citenamefont {Edwards}\ and\ \citenamefont
		{Wilkinson}(1982)}]{Edwards1982}%
	\BibitemOpen
	\bibfield  {author} {\bibinfo {author} {\bibfnamefont {S.~F.}\ \bibnamefont
			{Edwards}}\ and\ \bibinfo {author} {\bibfnamefont {D.~R.}\ \bibnamefont
			{Wilkinson}},\ }\href@noop {} {\bibfield  {journal} {\bibinfo  {journal}
			{Proc. R. Soc. London, Ser. A}\ }\textbf {\bibinfo {volume} {381}},\ \bibinfo
		{pages} {17} (\bibinfo {year} {1982})}\BibitemShut {NoStop}%
	\bibitem [{\citenamefont {Pang}\ and\ \citenamefont {Tzeng}(2004)}]{Pang2004}%
	\BibitemOpen
	\bibfield  {author} {\bibinfo {author} {\bibfnamefont {N.-N.}\ \bibnamefont
			{Pang}}\ and\ \bibinfo {author} {\bibfnamefont {W.-J.}\ \bibnamefont
			{Tzeng}},\ }\href {\doibase 10.1103/PhysRevE.70.011105} {\bibfield  {journal}
		{\bibinfo  {journal} {Phys. Rev. E}\ }\textbf {\bibinfo {volume} {70}},\
		\bibinfo {pages} {011105} (\bibinfo {year} {2004})}\BibitemShut {NoStop}%
	\bibitem [{\citenamefont {Olver}\ \emph {et~al.}(2010)\citenamefont {Olver},
		\citenamefont {Lozier}, \citenamefont {Boisvert},\ and\ \citenamefont
		{Clark}}]{Olver2010}%
	\BibitemOpen
	\bibinfo {editor} {\bibfnamefont {F.~W.~J.}\ \bibnamefont {Olver}}, \bibinfo
	{editor} {\bibfnamefont {D.~W.}\ \bibnamefont {Lozier}}, \bibinfo {editor}
	{\bibfnamefont {R.~F.}\ \bibnamefont {Boisvert}}, \ and\ \bibinfo {editor}
	{\bibfnamefont {C.~W.}\ \bibnamefont {Clark}},\ eds.,\ \href
	{https://dlmf.nist.gov/} {\emph {\bibinfo {title} {NIST handbook of
				mathematical functions}}}\ (\bibinfo  {publisher} {Cambridge University
		Press},\ \bibinfo {year} {2010})\BibitemShut {NoStop}%
	\bibitem [{\citenamefont {L{\'o}pez}(1999)}]{Lopez1999}%
	\BibitemOpen
	\bibfield  {author} {\bibinfo {author} {\bibfnamefont {J.~M.}\ \bibnamefont
			{L{\'o}pez}},\ }\href@noop {} {\bibfield  {journal} {\bibinfo  {journal}
			{Phys. Rev. Lett.}\ }\textbf {\bibinfo {volume} {83}},\ \bibinfo {pages}
		{4594} (\bibinfo {year} {1999})}\BibitemShut {NoStop}%
	\bibitem [{\citenamefont {Milstein}(1994)}]{Milstein1994}%
	\BibitemOpen
	\bibfield  {author} {\bibinfo {author} {\bibfnamefont {G.~N.}\ \bibnamefont
			{Milstein}},\ }\href@noop {} {\emph {\bibinfo {title} {Numerical integration
				of stochastic differential equations}}},\ Vol.\ \bibinfo {volume} {313}\
	(\bibinfo  {publisher} {Springer Science \& Business Media},\ \bibinfo {year}
	{1994})\BibitemShut {NoStop}%
	\bibitem [{\citenamefont {Mandelbrot}(1971)}]{Mandelbrot1971}%
	\BibitemOpen
	\bibfield  {author} {\bibinfo {author} {\bibfnamefont {B.~B.}\ \bibnamefont
			{Mandelbrot}},\ }\href
	{https://agupubs.onlinelibrary.wiley.com/doi/abs/10.1029/WR007i003p00543}
	{\bibfield  {journal} {\bibinfo  {journal} {Water Resour. Res.}\ }\textbf
		{\bibinfo {volume} {7}},\ \bibinfo {pages} {543} (\bibinfo {year}
		{1971})}\BibitemShut {NoStop}%
	\bibitem [{\citenamefont {Lam}\ \emph {et~al.}(1992)\citenamefont {Lam},
		\citenamefont {Sander},\ and\ \citenamefont {Wolf}}]{Lam1992}%
	\BibitemOpen
	\bibfield  {author} {\bibinfo {author} {\bibfnamefont {C.-H.}\ \bibnamefont
			{Lam}}, \bibinfo {author} {\bibfnamefont {L.~M.}\ \bibnamefont {Sander}}, \
		and\ \bibinfo {author} {\bibfnamefont {D.~E.}\ \bibnamefont {Wolf}},\ }\href
	{\doibase 10.1103/PhysRevA.46.R6128} {\bibfield  {journal} {\bibinfo
			{journal} {Phys. Rev. A}\ }\textbf {\bibinfo {volume} {46}},\ \bibinfo
		{pages} {R6128} (\bibinfo {year} {1992})}\BibitemShut {NoStop}%
\end{thebibliography}
%
\end{document}